\documentclass{PoS}
\usepackage{amsmath,amssymb,fontenc,times,mathptmx,graphicx}
\title{Recent results on Bose-Einstein correlations by the PHENIX Experiment}

\ShortTitle{Recent results on Bose-Einstein correlations by the PHENIX Experiment}

\author{\speaker{M\'at\'e Csan\'ad} for the PHENIX Collaboration\\
        E\"otv\"os University\\
        E-mail: \email{csanad@elte.hu}}

\abstract{Bose-Einstein momentum correlation functions of identical bosons reveal the shape
and size of the (soft) particle emitting source of the given particle. The widths of these 
correlation functions are called HBT radii, named after Brown and Twiss who studied the 
angular diameter of stars via intensity correlations in their radio telescopes. Today, high 
energy physics experiments measure the HBT radii as a function of many parameters: particle 
type, transverse momentum, azimuthal angle, collision energy, collision geometry. In this 
paper we present results from the RHIC PHENIX experiment. These include the observation of 
strong azimuthal-angle dependence of the extracted Gaussian HBT radii, the similarities and 
differences between kaon and pion HBT radii. The key point of this paper is the application 
of Bose-Einstein correlations to the search for the critical point: how HBT radii would show 
the appearance of a first order phase transition, and what the non-monotonic collision energy 
dependence of the pion source tells us about the critical point; and how the non-Gaussian 
shape of correlation functions is related to one of the critical exponents. }

\FullConference{9th International Workshop on Critical Point and Onset of Deconfinement\\
		 17-21 November, 2014\\
		 ZiF (Center of Interdisciplinary Research), University of Bielefeld, Germany}

\begin{document}

\section{Introduction}
Ultra-relativistic collisions of nuclei, so called ``Little Bangs'' are observed at the 
Relativistic Heavy Ion Collider (RHIC) of the Brookhaven National Laboratory, New York. The 
PHENIX Experiment at RHIC detects and identifies a large variety of the outcoming particles 
over a broad energy range. PHENIX has been participating  in the Beam Energy Scan program of 
RHIC, and collision data of many nuclei (d, He, Cu, Au, U) were taken at several center of 
mass energies (per nucleon), from 7.7 GeV to 200 GeV. The goal behind this variety of energies is to 
investigate the phase structure of QCD matter and the nature of the quark-hadron transition. 
This transition is a cross-over at low baryochemical potential~\cite{Aoki:2006we}, 
corresponding to top RHIC energy. It is however believed to be a first order phase transition 
at higher baryochemical potential (i.e. lower collision energies), while there may be a 
critical endpoint (CEP) somewhere in between. Correlation and fluctuation
measurements~\cite{Mitchell:2009zz,Lacey:2014rxa} are a powerful tool to be utilized in the 
search for direct evidence of the phase transition and the location of the critical point.  
In the present paper we focus on the soft pion production in these collisions, in particular 
the measurement of Bose-Einstein correlations between these low momentum pions. We discuss 
general features of Bose-Einstein correlations below, and then show results of various
Bose-Einstein correlation measurements.

\section{Basics of Bose-Einstein correlations and HBT radii}
R. H. Brown and R. Q. Twiss, radio astronomers, used two radio antennas pointed at a distant 
star and chose to detect each signal separately to take away its high frequency component. 
They saw an interference term present in the product of these two detected signals, and it 
turned out, that this interference enabled them to measure the angular radius of the observed 
star~\cite{HanburyBrown:1956pf}. This technique was then translated to high energy physics by 
Goldhaber et al.~\cite{Goldhaber:1960sf}, and it turned out, that Bose-Einstein correlations 
of pion pairs coming out of high energy nuclear collisions can be used to study the space-
time extent and shape of the particle emitting source.

The basic idea can be described in a simple quantum-mechanical picture: the detected two-
particle momentum distribution depends on the two-particle probability amplitude, i.e. the 
absolute square of the two-particle wave function. This wave function has to be symmetric for 
the exchange of the two particles, and in the end this results in a simple connection between 
the normalized particle emitting source $S(r)$ and the momentum-correlation function $C(q)$:
\begin{align}
C(q) = 1+ \left|\tilde S(q)\right|^2
\end{align}
where $\tilde S(q)$ is the Fourier transform of $S(r)$, and $q=p_1-p_2$ is the four-
momentum difference of the two identical bosons. In case of usual sources, the $C(q)
\rightarrow 1$ for $q \rightarrow \infty$, and $C(0)=2$ (as $\tilde S(0) = 1$
for normalized sources).

There are however many complications that distort the above simple picture. One of these is 
the Coulomb-interaction (and other final-state interactions, which however can be neglected 
in case of pions) between the two particles. This can be handled by using the solution of the 
two-particle Schrödinger-equation with a Coulomb-potential, as discussed in
Ref.~\cite{Alt:1998nr} for details. Another complication is, that a significant fraction of 
the produced pions comes from the decays of long lived resonances. These pions are produced 
very far from the hot, hydrodynamically expanding core, and give a very narrow contribution 
to $C(q)$. This narrow component cannot be resolved experimentally, thus the correlation 
function seems to have an intercept smaller than 2. Such a two-component source can be 
described in the core-halo picture (introduced in Refs.~\cite{Csorgo:1994in,Bolz:1992hc}), 
where the experimentally resolvable correlation function is
\begin{align}
C(q) = 1+ \lambda \left|\tilde S_{\rm core}(q)\right|^2
\end{align}
where $S_{\rm core}$ is the normalized source function of primordial pions, $0 < \lambda < 
1$, and $C(q\rightarrow 0)=1+\lambda$ for the experimentally resolvable $q$ region.

Usually, the 1D correlation function is measured as a function of $q_{\rm inv}=\sqrt{-q^2}$, 
but 3D information on the source can also be extracted from measurements, if statistics 
permits it. The $z$ or longitudinal (\emph{long}) direction is that of the beam. The plane 
perpendicular to it is parametrized with a pair-coordinate system introduced in
Refs.~\cite{Bertsch:1989vn,Pratt:1986cc}. In this system the \emph{out} direction is that of 
the average transverse momentum of the pair, while the \emph{side} direction is perpendicular 
to both \emph{out} and \emph{long}. Also, instead of the center of mass or the lab system, 
the longitudinally comoving system (LCMS) is used usually (introduced in
Ref.~\cite{Pratt:1990zq}). In a Bertsch-Pratt LCMS system, the average momentum is
$K^\mu = (M_t,K_t,0,0)$, while from the mass-shell condition one obtains
\begin{align}
q_0 = \frac{K_t}{M_t} q_{\rm out} = \beta_t q_{\rm out}
\end{align}
i.e. the time component of the momentum difference is related to the difference in the 
outward direction.

The main physics information is to be extracted from the shape of the measured correlation 
functions in one dimension (as a function of $q_{\rm inv}$) or in three dimensions (as a 
function of $q_{\rm out}$, $q_{\rm side}$ and $q_{\rm long}$).
The simplest case is to measure the width' of the correlation function (the so-called HBT 
radii named after H. Brown and Twiss), with a Gaussian assumption. This is not always valid 
however, and in an ideal case, model independent shape analysis has to be
pursued~\cite{Csorgo:1999wx}. Others restore the source image from the correlation function, 
via the method of imaging~\cite{Brown:2005ze}. In this paper we report about results of 
ordinary Gaussian analyses, and the measurement of Gaussian HBT radii.

And how are HBT radii connected to the the critical point or phase transitions?
It turns out, that in a simple hydrodynamic picture, the out and
side directed HBT radii are~\cite{Csorgo:1995bi}
\begin{align}
R_{\rm out}^2 = \frac{R^2}{1+\frac{M_t}{T_0}u_t^2}+\beta_t^2\Delta\tau^2
\qquad\textnormal{ and }\qquad
R_{\rm side}^2 = \frac{R^2}{1+\frac{M_t}{T_0}u_t^2},
\end{align}
which implies two things: radii scale as $1/\sqrt{m_t}$ with the average transverse mass of 
the pair, and out minus side difference is related to the emission duration $\Delta \tau$. 
This is particularly interesting, since a strong first order phase transition would cause a 
long emission and thus a large difference of these radii -- which was not the case 
experimentally~\cite{Adler:2004rq}. Readers shall however be reminded, that there is rich 
physics hidden in this story, and are encouraged to read Ref.~\cite{Pratt:2008qv} for an
in-depth analysis.

\section{PHENIX HBT results}
\begin{figure}
  \begin{center}
  \includegraphics[width=0.6\linewidth]{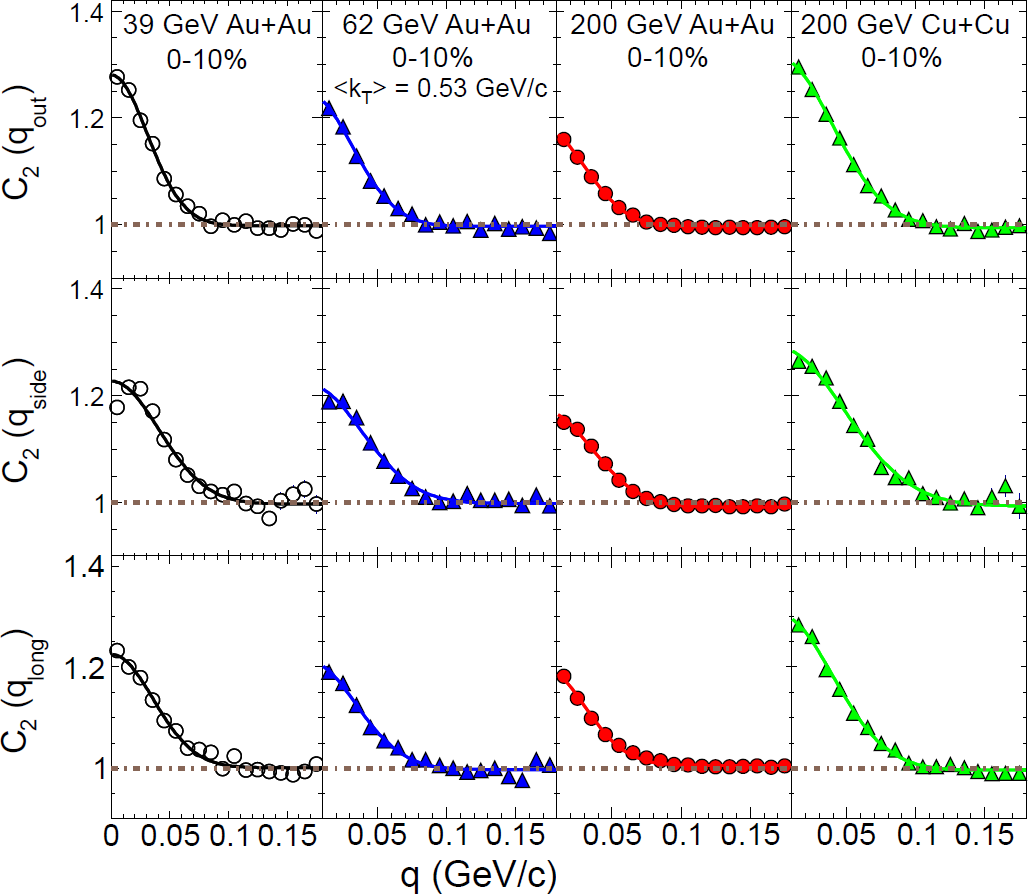}
\caption{A variety of three dimensional correlation functions at a glance, from Ref.~\cite{Adare:2014qvs}.
\label{f:corrall} }
  \end{center}
\end{figure}

In this section, we present recent PHENIX results on Bose-Einstein correlations. We show a 
few example correlation functions in Fig.~\ref{f:corrall}, from Ref.~\cite{Adare:2014qvs}. 
These (and others) were then analyzed by extracting their Gaussian width, i.e. the HBT radii. 
In the following subsections, we present such results on HBT radii.

\subsection{Transverse momentum scaling}

\begin{figure}
  \begin{center}
  \includegraphics[width=0.7\linewidth]{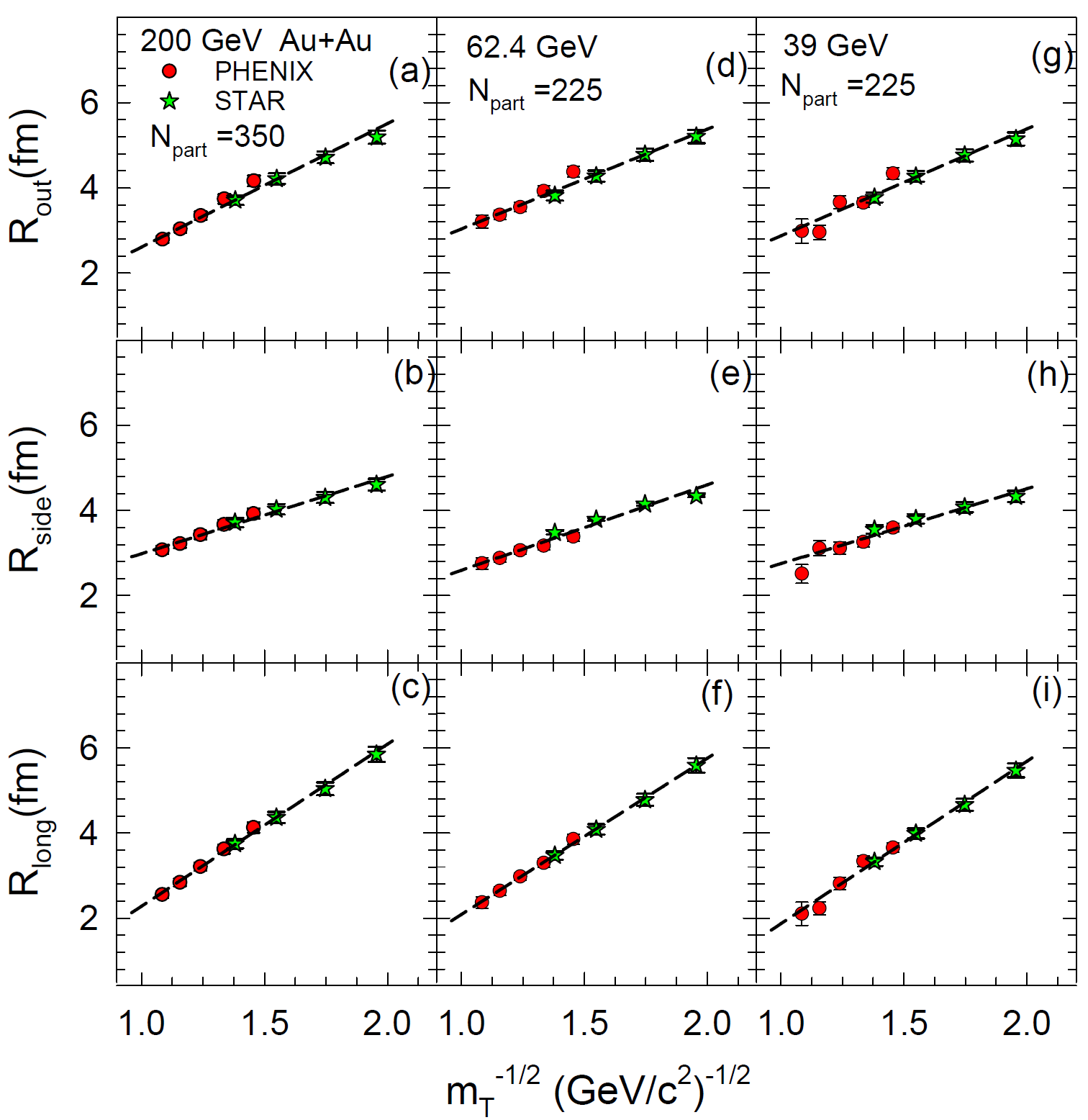}
\caption{Transverse momentum scaling of HBT radii from 39 to 200 GeV, from Ref.~\cite{Adare:2014qvs}.
\label{f:mtscaling} }
  \end{center}
\end{figure}

Correlation functions of pion pairs of several average transverse mass bins 
were measured in 39 to 200 GeV Au+Au collisions~\cite{Adare:2014qvs}. The $m_t$
dependence of the resulting Gaussian radii $R_{\rm out}$, $R_{\rm side}$, and $R_{\rm long}$,
compared to recent STAR data~\cite{Adamczyk:2014mxp} are shown in Fig.~\ref{f:mtscaling}.
The data of different centrality selections show characteristic $1/\sqrt{m_t}$ 
scaling patterns as a function of the transverse mass $m_t$
of the emitted pion pairs, consistent with hydrodynamic expansion.
A good level of agreement is seen between the PHENIX and STAR
data sets, while the PHENIX data extends the covered $m_t$ range
of RHIC HBT radii. It is important to note that the scaling also holds in 200 GeV
d+Au collisions~\cite{Mwai:2014pda}.

\begin{figure}
  \begin{center}
  \includegraphics[width=0.75\linewidth]{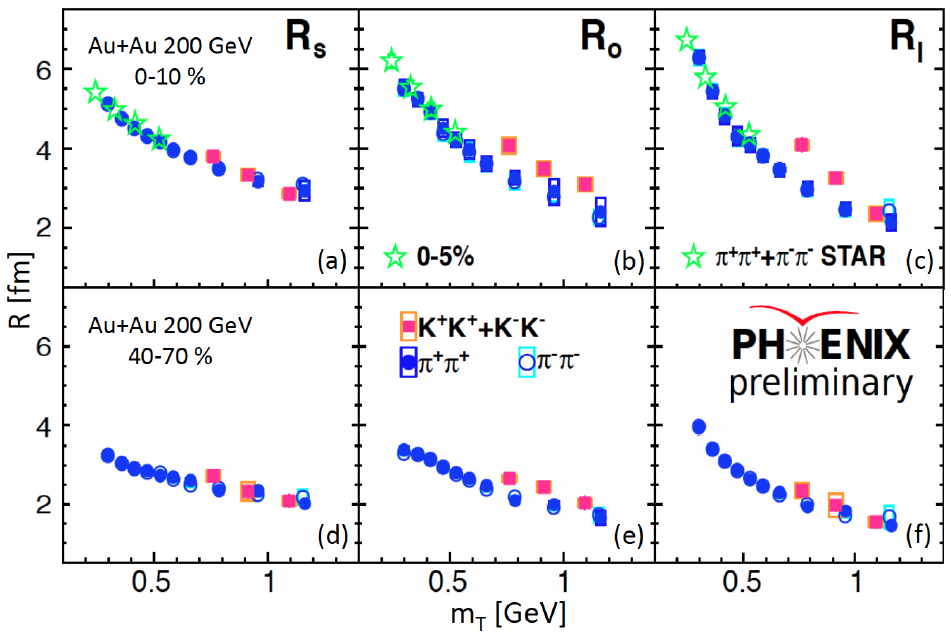}
\caption{Transverse momentum scaling of kaon radii, from Ref.~\cite{Niida:2013dza}.
Significant difference from pion radii is observed in central collisions, in the out and
longitudinal directions.
\label{f:kaonmtscaling} }
  \end{center}
\end{figure}

Correlation functions of kaon pairs were also recently measured at PHENIX. Since the pion
sample is more affected by hadronic rescattering and the decays of long lived resonances,
kaons represent a more clean proble of the particle emitting source. PHENIX recently
measured kaon HBT radii at several centralities of 200 GeV Au+Au
collisions~\cite{Niida:2013dza}. Fig.~\ref{f:kaonmtscaling} shows a comparison of
the $m_t$ dependence of HBT radii between charged pions and kaons. There is no significant
difference in the side direction (i.e. $m_t$ scaling of different particle species holds),
while the kaon radii in the out and longitudinal directions are slightly larger than the
pions (at the same $m_t$). The difference is more pronounced in central collisions.
These results may point to different dynamics and duration of kaon production as
compared to pion production.

\subsection{Azimuthally dependent HBT}
\begin{figure}
  \begin{center}
  \includegraphics[width=0.9\linewidth]{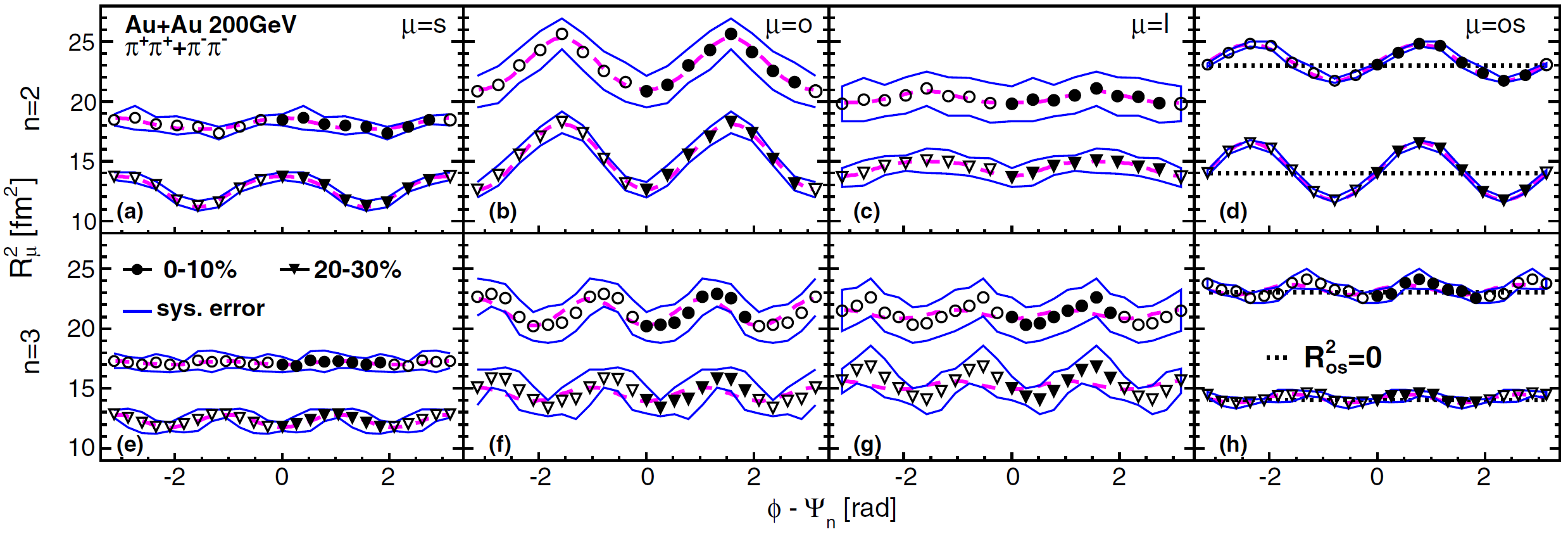}
  \includegraphics[width=0.55\linewidth]{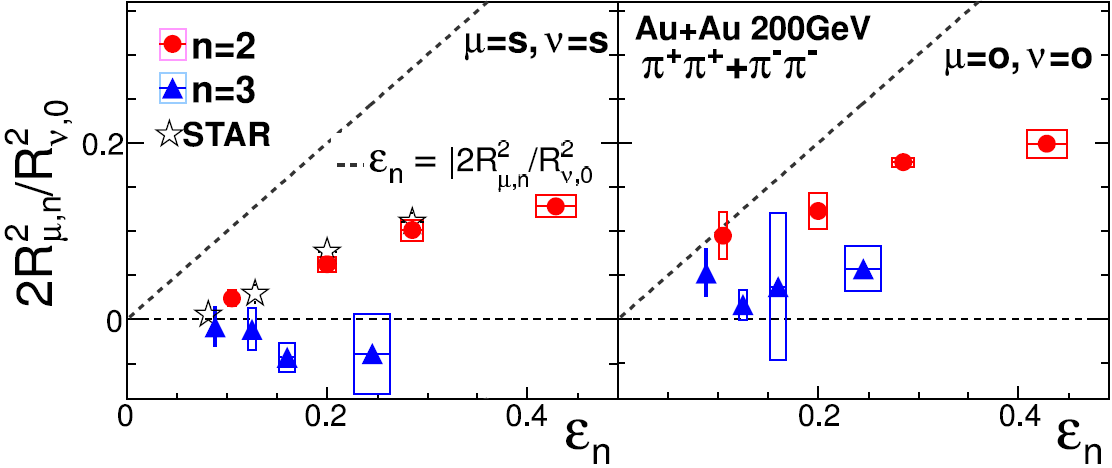}
\caption{Azimuthally sensitive HBT radii w.r.t. the 2nd and 3rd order event planes (top panel),
from Ref.~\cite{Adare:2014vax}. Their first Fourier-components as a function of initial eccentricity
are shown in the bottom panel.\label{f:ashbt} }
  \end{center}
\end{figure}

HBT radii were recently measured at PHENIX with respect to the 2nd- and 3rd-order event plane
in 200 GeV Au+Au collisions~\cite{Adare:2014vax}. A strong azimuthal-angle dependence of the
Gaussian HBT radii was observed with respect to both the 2nd- and 3rd-order event planes,
as shown in Fig.~\ref{f:ashbt} (top panel). The oscillation amplitudes were extracted by fitting
$R_{\mu,0}^2+2\sum\limits_{n=m,2m}R_{\mu,n}^2\cos[n(\phi-\Psi_m)]$ type of functional
forms. The amplitude ratios $2R_{\mu,n}^2/R_{\nu,0}^2$ versus initial spatial anisotropies
$\epsilon_n$ (from Monte-Carlo Glauber simulations) are shown also in Fig.~\ref{f:ashbt} (bottom
panel). Since these ratios are sensitive to the final eccentricity, the results for $n=2$ 
indicate that the initial eccentricity is reduced during the medium evolution,
but is not reversed. Results for $n=3$ however indicate, that the initial triangular
shape may be reversed by the end of the medium evolution, and that the 3rd-order oscillations
are largely dominated by the dynamical effects from triangular flow, as detailed in
Ref.~\cite{Adare:2014vax}

A new technique in measuring azimutally sensitive HBT radii is the application of event shape
selection. In this method one controls not just the direction of the event plane in each event,
but also the magnitude of the second order flow vector $Q_2$, measured by the Reaction
Plane Detector (RxNP). $Q_2$ is defined as $\sum_i w_i(\cos(2\phi_i),\sin(2\phi_i))$ with $w_i$
being the multiplicity in the $i$-th segment of RxNP.
The measured eccentricity (defined as $2R_{\mu,n}^2/R_{\nu,0}^2$) was found to be enhanced
for events with higher magnitude of the flow vector $Q_2$, as detailed in e.g. Ref.~\cite{Niida:2015ifa}.

\subsection{Geometric scaling}
\begin{figure}
  \begin{center}
  \includegraphics[width=0.8\linewidth]{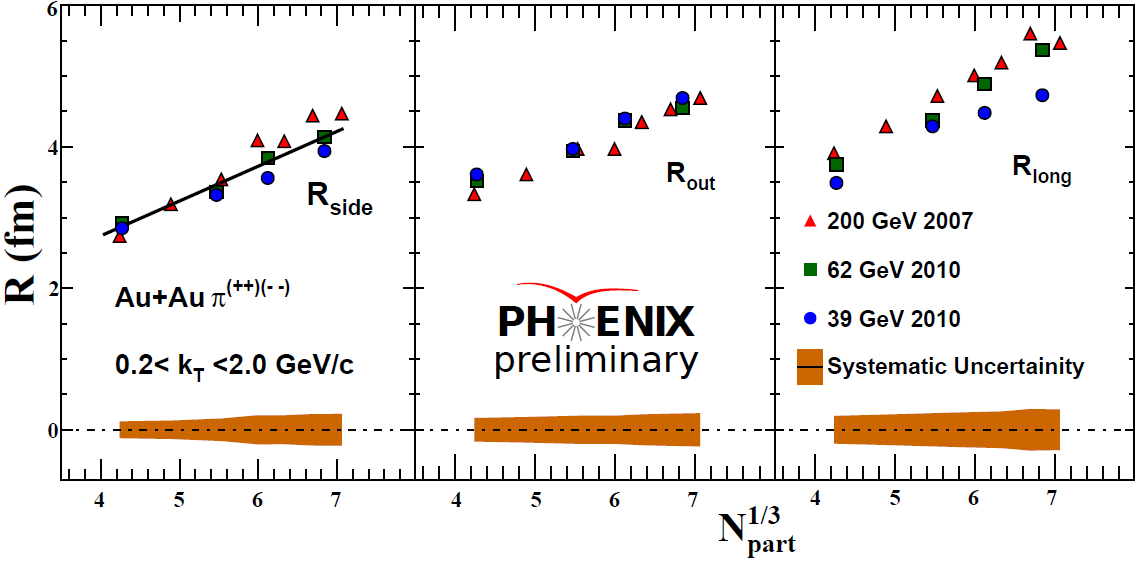}
  \includegraphics[width=0.8\linewidth]{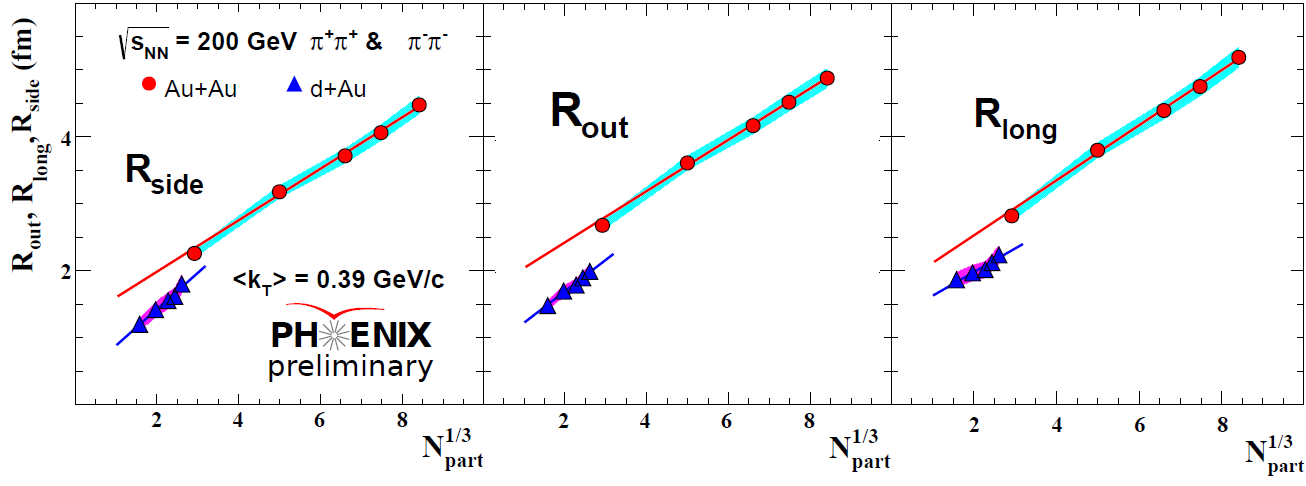}
\caption{System size scaling from Refs.~\cite{Mwai:2014phd,Mwai:2014pda}. Linear scaling
is observed, but d+Au data seem to fall on a different scaling curve.
\label{f:geomscaling} }
  \end{center}
\end{figure}

As the initial system volume scales with $N_{\rm part}$, the $N_{\rm part}^{1/3}$ dependence of the HBT
radii is often analyzed. Recently, HBT radii measurements in 39, 62 and 200 GeV Au+Au collisions were
compared~\cite{Mwai:2014phd}, as shown in Fig.~\ref{f:geomscaling} (top panel). All energies show
a linear dependence, approximately independently of the beam energy. This indicates that at these RHIC
energies, the dynamics of the system does not change dramatically, similar $N_{\rm part}$ means
similar initial system size, which in turn leads to a similar homogeneity length.

PHENIX also measured three dimensional HBT radii in d+Au collisions as a function of
centrality~\cite{Mwai:2014pda}. The linear $N_{\rm part}^{1/3}$ dependence shows a
similar slope in Au+Au and d+Au collisions (at the same average transverse momentum).
The similarity of the slope is stronger in case of $R_{\rm out}$ and $R_{\rm side}$ (hinting
at similar transverse dynamics of the two systems), while the $R_{\rm long}$ slope is slightly
different, maybe due to different longitudinal dynamics in d+Au and Au+Au collisions. Results
are shown in Fig.~\ref{f:geomscaling} (bottom panel).

\begin{figure}
  \begin{center}
  \includegraphics[width=0.9\linewidth]{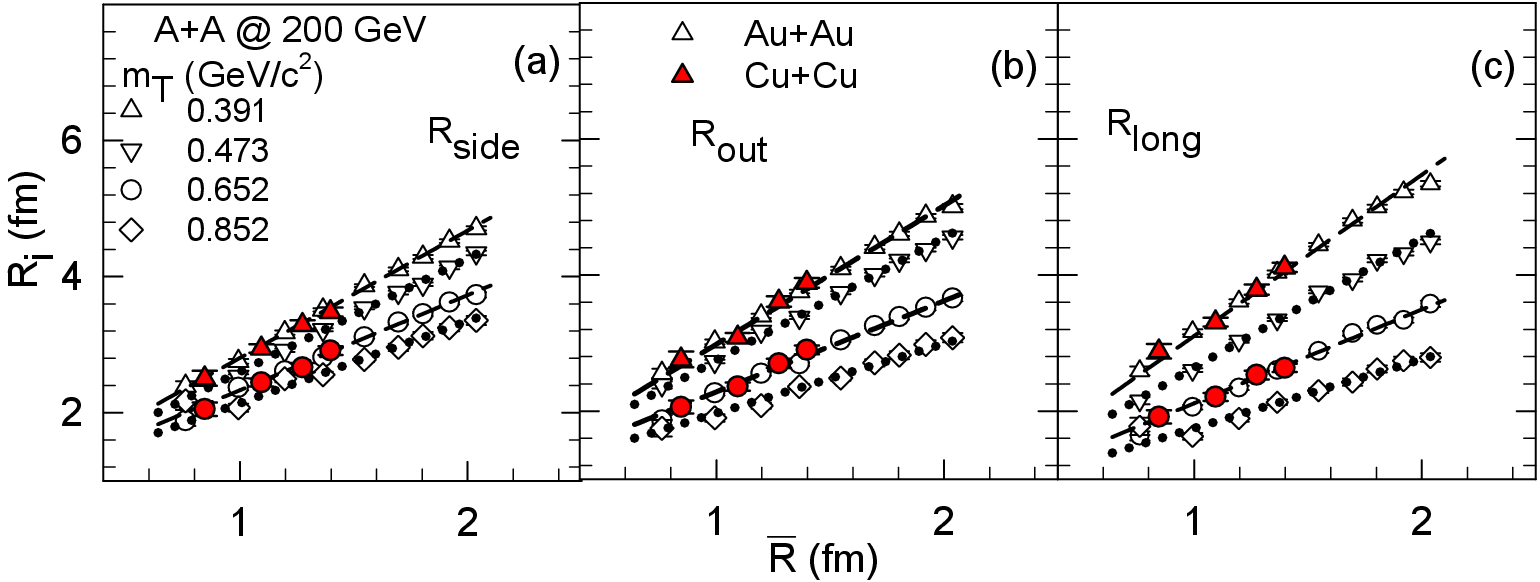}
\caption{Geometric scaling versus $\overline{R}$, from Ref.~\cite{Adare:2014qvs}.
\label{f:rbarscaling} }
  \end{center}
\end{figure}

Besides the above system size scaling, more detailed geometrical scalings were discovered.
These are based on the characteristic initial transverse size, defined as
$1/\overline{R}^2 = 1/\sigma_x^2+1/\sigma_y^2$
with $\sigma_{x,y}$ being the RMS widths of participant nucleon distributions
in the $x$ and $y$ directions (estimated via Glauber Monte Carlo calculations).
The space-time extent of the emission source at freeze-out, as measured by the
HBT radii, reflects the initial size of the system, the expansion time or lifetime $\tau$,
as well as transverse expansion rate (visible via the $1/\sqrt{m_t}$ dependence
of the HBT radii, as discussed in Section.~2). Since the expansion time is expected
to scale with $\overline{R}$, the HBT radii are also expected to scale with $\overline{R}$
in a given $m_t$ range. This scaling was indeed observed at PHENIX~\cite{Adare:2014qvs}
as shown in Fig.~\ref{f:rbarscaling}. Note furthermore, that it also holds in case of d+Au
collisions~\cite{Mwai:2014pda}.
The detailed scaling laws indicate that $\overline{R}$
is a better predictor of final state homogeneity length, than $N_{\rm part}^{1/3}$,
which is consistent with observations that HBT is sensitive to the expansion dynamics.

\subsection{Excitation function}
\begin{figure}
  \begin{center}
  \includegraphics[width=0.6\linewidth]{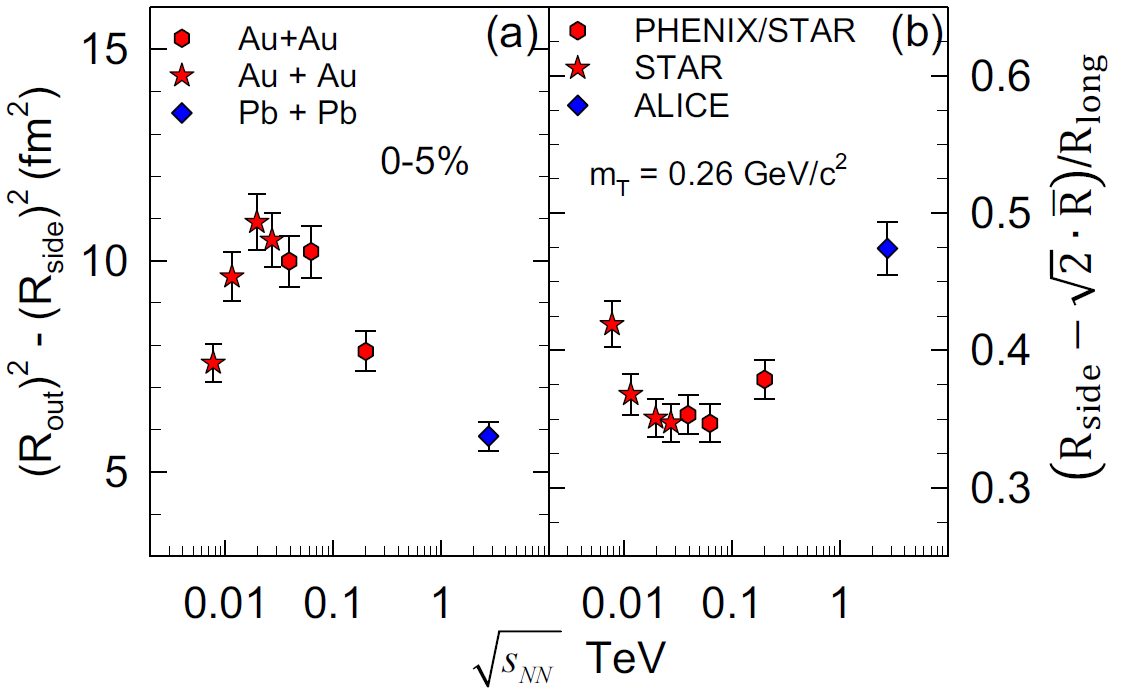}
\caption{Excitation function of different combinations of HBT radii, from Ref.~\cite{Adare:2014qvs}.
Non-monotonicity is seen around $\sqrt{s_{\rm NN}}=$40 GeV.
\label{f:excitfunc} }
  \end{center}
\end{figure}

In order to access physical descriptors of the expanding medium
like emission duration and expansion velocity, quantities
$R_{\rm out}^2-R_{\rm side}^2$ and
$(R_{\rm side}-\sqrt{2}\overline{R})/R_{\rm long}$ were measured,
as discussed in Ref.~\cite{Adare:2014qvs}. These were investigated
as a function of collision energy, and results at $m_t=0.26$ GeV$/c^2$,
for 0-5\% centrality are shown in Fig.~\ref{f:excitfunc}. The observed
non-monotonic pattern near $\sqrt{s_{\rm NN}}=$40 GeV could be an indication of the vicinity of the softest
point of the QCD EoS. It is however important to note that the 
measured HBT radii are smaller than the actual sizes of the system,
and are linked to expansion dynamics. Due to the amount and depth of
physics involved in these measurements, further experimental studies and
full hybrid models are required to confirm if the observed non-monotonic 
patterns are linked to the critical end point in the QCD phase diagram.

\subsection{Levy exponents}
At the critical point universal power law behavior of several observables
can be seen as a function of the temperature of the system, characterized
by critical exponents, one being that of the spatial correlation function. Levy
distributions are present in the experimental momentum correlation functions
and the spatial correlation functions restored from it. The Levy-exponent $\alpha$ of
these correlation functions is equivalent to the spatial correlation exponent,
conjectured to be near 0.5 in case of the critical end point~\cite{Csorgo:2003uv}.
PHENIX performed measurement of the Levy-exponent in 200 GeV Au+Au
collisions~\cite{Csanad:2005nr}. At this energy, the exponent is far from
both the Gaussian case ($\alpha=2$) and  the critical case ($\alpha=0.5$).
A detailed measurement of the excitation function of the correlation exponent
is expected to substantially refine results on the QCD critical end point.

\section{Beyond HBT}
As present conference volume is about the search for and characterization of
the QCD critical end point, this paper cannot be complete without mentioning
some more related PHENIX results, beyond HBT.

One set of important results is that of neutral pion spectra, which were measured in 39 GeV
to 200 GeV Au+Au collisions. Midrapidity nuclear modification factors measured in the
$1<p_t<10$ GeV$/c$ range were compared to each other in Ref.~\cite{Adare:2012uk}. 
For central collisions, suppression is significant at $\sqrt{s_{\rm NN}}=$39, 62 and
200 GeV as well. On the other hand, in mid-peripheral 39 GeV collisions, the nuclear
modification factor $R_{\rm AA}$ is consistent with unity above $p_t>3$ GeV$/c$.
  
PHENIX recently measured the anisotropic flow coefficients $v_n$ for identified pions,
kaons and protons, relative to the $n$-th order event planes, in 200 GeV Au+Au
collisions~\cite{Adare:2014kci}. These flow coefficients show characteristic patterns
consistent with hydrodynamical expansion of the matter produced in the collisions.
They all fall on modified quark number scaling curves, if $v_n/(n_q)^{n/2}$ is
plotted versus ${\rm KE}_T/n_q$. This may indicate a hydrodynamic
expansion, a quark coalescence process, and/or an acoustic nature of the QGP.
Interestingly, these scalings with the number of constituent quarks hold for $n=2,3$
also in 39 and 62 GeV Au+Au collisions~\cite{Gu:2012br}.

\section{Summary}
A wide variety of HBT data is now available. Geometric and hydrodynamic
scalings are observed for a broad $\sqrt{s_{\rm NN}}$ range: HBT
radii versus $1/\sqrt{m_t}$, versus $N_{\rm part}^{1/3}$ and versus
$\overline{R}$ follow simple linear curves. Excitation function of HBT
radii however show a non-monotonic behavior near $\sqrt{s_{\rm NN}}=40$
GeV. This may be attributed to non-monotonicity in the emission duration;
more detailed analysis and measurements have however to be performed
to understand and confirm the relation of this behavior to the QCD 
critical end point. Measurements of the correlation exponent could
for example substantially refine our understanding. There are however
other probes measured in the vicinity of this point. Non-monotonic
behavior is seen in neutral pion nuclear modifications, while quark
number scaling is preserved from 200 GeV to 39 GeV. All of these
observables and their excitation functions shuld be understood in a
global complex picture, in order to acquire detailed knowledge of the
QCD critical end point.

\bibliographystyle{../../../utphys}
\bibliography{../../../Master}

\end{document}